\newcommand{\beq}{\begin{equation}}
\newcommand{\eeq}{\end{equation}}
\newcommand{\bea}{\begin{eqnarray}}
\newcommand{\eea}{\end{eqnarray}}
\newcommand{\bear}{\begin{eqnarray*}}
\newcommand{\eear}{\end{eqnarray*}}
\newcommand{\lb}{\label}
\newcommand{\rf}[1]{(\ref{#1})}
\documentstyle[aps,preprint,doublespace]{revtex}

\begin{document}

\draft

\title
{The Exact Solution of the Asymmetric Exclusion Problem With Particles of Arbitrary Size: Matrix Product {\it Ansatz}}

\author{Francisco C. Alcaraz and Matheus J. Lazo }

\address{Universidade de S\~ao Paulo, Instituto de F\'{\i}sica de S\~ao Carlos, Caixa Postal 369, 13560-590,\\ S\~ao Carlos, SP,
Brazil}

\maketitle

\begin{abstract}

        The exact solution of the asymmetric exclusion problem and several of its generalizations is obtained by a matrix product {\it ansatz}. Due to the similarity of the master equation and the Schr\"odinger equation at imaginary times the solution of these problems reduces to the diagonalization of a one dimensional quantum Hamiltonian. We present initially the solution of the problem when an arbitrary mixture of molecules, each of then having an arbitrary size ($s=0,1,2, \ldots$) in units of lattice spacing, diffuses asymmetrically on the lattice. The solution of the more general problem where we have  the diffusion of particles belonging to $N$ distinct class of particles ($c=1, \ldots ,N$), with hierarchical order, and arbitrary sizes is also solved. Our matrix product {\it ansatz} asserts that the amplitudes of an arbitrary eigenfunction of the associated quantum Hamiltonian can be expressed by a product of matrices. The algebraic properties of the matrices defining the {\it ansatz} depend on the particular associated Hamiltonian. The absence of contradictions in the algebraic relations defining the algebra ensures the exact integrability of the model. In the case of particles distributed in $N>2$ classes, the associativity of the above algebra implies the Yang-Baxter relations of the exact integrable model.

\end{abstract}
\vspace{0.5cm}

\pacs{
}

\narrowtext  
\section{Introduction}

The representation of interacting stochastic particle dynamics in terms of quantum spin systems produced interesting and fruitful interchanges among the area of equilibrium and nonequilibrium statistical mechanics. The connection among these areas follows from the similarity between the master equation describing the time-fluctuations on the nonequilibrium stochastic problem and the quantum fluctuations of the equilibrium quantum spin chains~\cite{lushi}-\cite{schu-domb}.

        Unlike the area of nonequilibrium interacting systems, where very few models are fully solvable, there exists a huge family of quantum chains appearing in equilibrium problems that are exactly integrable. The machinery that allows the exact solutions of these quantum chains comes from the Bethe ansatz on its several formulations (see~\cite{baxter}-\cite{revschlo} for reviews). The above mentioned mathematical connection between equilibrium and nonequilibrium revealed that some quantum chains related to interacting stochastic problems are exactly solvable through the Bethe {\it ansatz}. The simplest example is the problem of asymmetric diffusion of hard-core particles on the one dimensional lattice (see~\cite{der3,ligget,schu-domb} for reviews). The time fluctuations of this last model is governed by a time evolution operator that coincides with the exact integrable anisotropic Heisenberg chain, or the so called, XXZ quantum chain, in its ferromagnetically ordered regime. A generalization of this stochastic problem where exact integrability is also known~\cite{bold}-\cite{ferra2} is the case where there exists $N$ ($N=1,2,...$) classes of particles hierarchically ordered diffusing asymmetrically on the lattice. The quantum chain related to this problem is known in the literature as the anisotropic Sutherland model~\cite{sutherland} or SU($3$) Perk-Schultz model~\cite{perkshultz}. In~\cite{PRE},~\cite{BJP1} and~\cite{BJP2} it was shown that the above mentioned asymmetric exclusion problem could also be solved exactly though the Bethe {\it ansatz} in the cases where the particles diffusing on the lattice have hard-core interactions of arbitrary range, or equivalently, the particles have size $s=0,1,2,...$, in units of lattice spacing.

        On the other hand, along the last decade it has been shown that the stationary distribution of probability densities of some stochastic models can also be expressed in terms of a matrix product {\it ansatz}. This means that the ground state eigenvector of the related quantum chain is also given by a matrix product {\it ansatz}. According to this {\it ansatz} the components of the ground state wavefunction are given in terms of a product of matrices. These components, apart from an overall normalization constant, are fixed by the commutation relations of the matrices defining the matrix product {\it ansatz}. These models are in general not exact integrable~\cite{dasmar} and the matrix product {\it ansatz} only gives the ground state wavefunction of the related Hamiltonian. Despite of this limitation this {\it ansatz} produced interesting results in a quite variety of problems including interface growth~\cite{krug}, boundary induce phase transitions~\cite{derr1}-\cite{derrevans}, the dynamics of shocks~\cite{derlebov} or traffic flow~\cite{nagel}. 

        An interesting development of the matrix product {\it ansatz} that happened also in the area of interacting stochastic models is nowadays known as the dynamical matrix {\it ansatz}~\cite{stinchshutz}. According to this new {\it ansatz}, whenever it is valid, the probability density of the stochastic system is given by a matrix product {\it ansatz} not only at the stationary state but at arbitrary times. In the related quantum chain this would be equivalent to the requirement that not only the ground state wavefunction, but an arbitrary one, should have its components given by a matrix product {\it ansatz}. The dynamical matrix product {\it ansatz} was applied originally on the problem of asymmetric diffusion of particles on the lattice~\cite{stinchshutz,sasamowada1}. More recently~\cite{popkov},~\cite{popkov2} it was also shown  that this {\it ansatz} can also be formulated in the problem of asymmetric diffusion of two types of particles. The validity of the {\it ansatz} was confirmed in the regions where the model is know to be exact integrable through the Bethe {\it ansatz}~\cite{alcrit1,BJP1}. Motivated by this fact we decided to verify if we can solve the above quantum chains directly though a matrix product {\it ansatz}, without considering any time dependence as happens in the dynamical matrix {\it ansatz}. Surprisingly we were able to rederive all the results previously obtained though the Bethe {\it ansatz} for the asymmetric diffusion problem with one specie of particles~\cite{PRE} or more~\cite{BJP1,BJP2}. Moreover our derivation turns out to be quite simple and it is not difficult to extend it to many other quantum Hamiltonians related or not to stochastic particle dynamics~\cite{TB}. We are going to present in this paper these derivations and as we shall see, many of the results obtained in~\cite{PRE},~\cite{BJP1} and~\cite{BJP2} can now be rederived quite easily. The simplicity of our {\it ansatz} enabled us to extend the results of~\cite{BJP2} to the case where each individual particle $i$ belonging to any class ($c=1,\ldots ,N$) is distinguishable with a given size $s_i$ ($s_i=0,1, \ldots$).

        The paper is organized as follows. In the next section we review the asymmetric diffusion problem with a single type of particles of arbitrary size and we derive the associated quantum chain. In section $3$ we introduce the matrix product {\it ansatz} and obtain the exact solution of the model presented in section $2$. Similarly as in section $2$, in section $4$, we derive the quantum Hamiltonian associated to the problem of asymmetric diffusion of several types of particles with arbitrary sizes and hierarchical order. In section $5$ the exact solution of the general model of section $4$ is obtained though an appropriate matrix product {\it ansatz}. Finally in section $6$ we conclude our paper with some final comments and conclusions.

\section{The asymmetric exclusion model with particles of arbitrary sizes}

        The standard asymmetric exclusion model is a one-dimensional stochastic model that describes the time fluctuations of hard-core particles diffusing asymmetrically on the lattice. If we denote an occupied site $i$ on the lattice by $\sigma_i^z=+1$ and a vacant site $i$ by $\sigma_i^z=-1$, the time evolution operator of the probability distribution of particles is given by the following asymmetric XXZ Hamiltonian
\beq
H=-\sum_{i=1}^{L}\left[\epsilon_{+}\sigma_{i}^{-}\sigma_{i+1}^{+}+\epsilon_{-}\sigma_{i}^{+}\sigma_{i+1}^{-}+
\frac{1}{4}(1-\sigma_{i}^{z}\sigma_{i+1}^{z})\right],
\lb{e1}
\eeq
where $L$ is the number of lattice sizes and $\sigma^{\pm}=(\sigma^{x}{\pm}i\sigma^{y})/2$ are the raising and lowering spin$-\frac{1}{2}$ Pauli operators. Periodic boundary conditions are imposed and $\epsilon_{+}$ and $\epsilon_{-}$ ($\epsilon_{+}+\epsilon_{-}=1$) are the transition probabilities for the motions to the right and left, respectively. It is important to notice that this Hamiltonian, contrary to the standard XXZ quantum chain, is not Hermitian for $\epsilon_+ \neq \epsilon_-$. Such property, besides producing complex value eigenvalues also produces massless regime, in a region where the standard XXZ is massive (gapped), whose mass gap vanishes as $L^{-3/2}$\cite{spohn}-\cite{kim2}.

        The generalization of this problem, that we consider in this section, is obtained by considering each distinct particle, instead of having an excluded hard-core volume of a single lattice size ($s=1$), may now have a hard-core volume of $s$ sites ($s=0,1,2,...$). Equivalently, each individual particle on the lattice may have a distinct size $s=0,1,2,...$. Particles of sizes $s$ on the lattice are composed by $s$ one-site monomerers and we represent their coordinates by giving the position of their leftmost monomer. In Fig. 1 some examples were shown for the configurations with $n=5$ molecules and some size distributions $\{s\}$ in a lattice with $L=5$ sites. We should notice that molecules of size $s=0$ has no excluded volume interaction and we can have an arbitrary number of them in a given site. However we should stress that although being sizeless they keep the order of the size's distribution on the lattice. This means that if a given particle of size $s$ is initially between particles of size $s'$ and $s''$ it will keep this relative order in future times.

        In order to describe the occupancy of a given site $i$ ($1,2,...,N$) we attach on it a site variable $\beta_i$ taking integer values ($\beta_i \in \bf{Z}$). If $\beta_i=0$, the site is vacant (or may be occupied by a monomer of the molecule on its leftmost neighboring site). If $\beta_i>0$, we have on the site a molecule of size $s=\beta_i$ and the sites $j=i+1,...,i+\beta_i -1$ are empty sites. Finally if $\beta_i=-n<0$ we have, at the site $i$, $n$ molecules of size zero. The allowed configurations, denoted by $\lbrace\beta_i\rbrace=\{ \beta_1,\beta_2,...,\beta_N\}$ are those satisfying the hard-core constraints imposed by the sizes of the molecules on the periodic lattice. This means that if in a given configuration $\lbrace\beta_i\rbrace$ we have $\beta_j\neq 0$ and $\beta_l\neq 0$ then we should have $l-j\geq \beta_l$ or $j-l\geq \beta_j$ depending if $l>j$ or $l<j$, respectively (see Fig. 1). 

        The master equation for the probability distribution at a given time $t$, $P(\lbrace\beta\rbrace,t)$, can be written in general as
\beq
\label{e2} 
\frac{\partial  P(\lbrace\beta\rbrace,t)}{\partial t} = -  \Gamma(\lbrace\beta\rbrace 
\rightarrow \lbrace\beta'\rbrace)  P(\lbrace\beta\rbrace,t) + \Gamma(\lbrace\beta'\rbrace 
\rightarrow \lbrace\beta\rbrace)  P(\lbrace\beta'\rbrace,t)
\eeq
where $\Gamma(\lbrace\beta\rbrace \rightarrow \lbrace\beta'\rbrace)$ is 
the transition rate where a configuration $\lbrace\beta\rbrace$ changes to $\lbrace\beta'\rbrace$. In the model we are considering there exists only diffusion processes. The allowed motions, whenever there is no hard-core constraints, are those in which a given particle diffuses to its right,
\bea
\label{e3}
\beta_i\;\;\emptyset_{i+1} &\rightarrow& \emptyset_i\;\;\beta_{i+1},\quad \beta > 0\nonumber\\
\beta_i\;\;\gamma_{i+1} &\rightarrow& (\beta+1)_i\;\;(\gamma-1)_{i+1},\quad \beta < 0,\quad \gamma \leq 0, 
\eea
with transition rate $\epsilon_R$, and diffusion to the left, 
\bea
\label{e4}
\emptyset_i\;\;\beta_{i+1} &\rightarrow& \beta_i\;\;\gamma_{i+1},\quad \beta > 0\nonumber\\
\gamma_i\;\;\beta_{i+1} &\rightarrow& (\gamma-1)_i\;\;(\beta+1)_{i+1},\quad \beta < 0,\quad \gamma \leq 0, 
\eea
with transition rate $\epsilon_L$. The master equation \rf{e2} can be written as a Schr\"odinger equation in Euclidean time (see~\cite{alcrit1} for general applications for two-body processes)
\beq 
\label{e5}
\frac{\partial  | P\rangle}{\partial t}=-H | P\rangle,
\eeq
if we interpret $| P\rangle \equiv P(\lbrace\beta\rbrace,t)$ as the associated wave function. If we represent $\beta_i$ as $|\beta\rangle_i$, the vectors $|\beta\rangle_1\otimes |\beta\rangle_2\otimes \cdots \otimes|\beta\rangle_N$ will span the associated Hilbert space. The diffusion process given in \rf{e3} and \rf{e4} give us the Hamiltonian~\cite{alcrit1}
\bea
\label{e6}
H &=& -D {\cal P} \sum_{i=1}^{L}\left( H_{i,i+1}^> + H_{i,i+1}^<  \right) {\cal P},\nonumber \\
H_{i,j}^> &=&  \sum_{\beta=1}^\infty \left[ \epsilon_+(E_{i}^{0, \beta}
E_{j}^{\beta, 0} - E_i^{\beta, \beta} E_{j}^{0, 0}) +
\epsilon_-(E_i^{\beta, 0}E_{j}^{0, \beta} - E_i^{0, 0} E_{j}^{\beta, \beta}) 
\right], \nonumber \\
H_{i,j}^< &=& \sum_{\beta=-\infty}^{-1} \sum_{\gamma=-\infty}^{0}\left[ \epsilon_+(E_{i}^{\beta+1, \beta}
E_{j}^{\gamma-1, \gamma} - E_i^{\beta, \beta} E_{j}^{\gamma, \gamma}) +
\epsilon_-(E_i^{\gamma-1, \gamma} E_{j}^{\beta+1, \beta} - E_i^{\gamma, \gamma} E_{j}^{\beta, \beta}) 
\right],
\eea
with 
\beq
\label{e7}
D = \epsilon_R + \epsilon_L,\;\;\; \epsilon_+ =\frac{\epsilon_R}{\epsilon_R + 
\epsilon_L},\;\;\; \epsilon_- = \frac{\epsilon_L}{\epsilon_R + \epsilon_L},
\eeq
and periodic boundary conditions. The matrices $E^{\alpha, \beta}$ are infinite-dimensional with a single nonzero element $(E^{\alpha, \beta})_{i, j} = \delta_{\alpha, i} \delta_{\beta, j}$ ($\alpha, \beta, i, j \in \bf{Z}$). The projector ${\cal P}$ projects out the configurations $|\{ \beta \} \rangle>$ satisfying the constraint that for all $\beta_i, \beta_j \neq 0 : (j-i)\geq s_i$ if $j>i$ or $(i-j)\geq s_j$ if $i>j$. The constant $D$ in \rf{e6} fixes the time scale and for simplicity we chose $D=1$. A simplification of our general problem happens when all the particles have the same size $s>0$. In this case the matrices $E^{\alpha, \beta}$ can be replaced by the spin-$\frac{1}{2}$ Pauli matrices and the Hamiltonian is given by
\beq
\label{e8}
H_{\{ s_1=\cdots =s_n=s\} }= -{\cal P}_s \left(\sum_{i=1}^L\left[\epsilon_+ \sigma_i^- \sigma_{i+1}^+ +\epsilon_- \sigma_i^+ \sigma_{i+1}^- \right] + \frac{1}{2}(\epsilon_+ + \epsilon_- ) (\sigma_i^z \sigma_{i+1}^z - 1)\right){\cal P}_s,
\eeq
where now ${\cal P}_s$ projects out the configuration where two up spins, in the $\sigma^z$-basis, are at distance smaller than the size $s>0$ of the particles. The simplest case $s=1$ gives ${\cal P}_s=1$ and we obtain the standard asymmetric exclusion Hamiltonian \rf{e1}. For the sake of comparison with the standard XXZ chain, normally considered in the context of magnetic systems, we perform for $\epsilon_+, \epsilon_- \neq 0$ the following canonical transformation:
\beq
\label{e9}
\sigma_i^{\pm} \rightarrow (\frac{\epsilon_-}{\epsilon_+})^{\pm \frac{i}{2}} \sigma_i^{\pm}, \sigma^z \rightarrow \sigma^z, i=1,2,\ldots ,L,
\eeq
in \rf{e8} and obtain
\bea
\label{e10}
H &=& -\frac{1}{2}\sqrt{\epsilon_+ \epsilon_-}\sum_{i=1}^L{\cal P}_s\left[\sigma_i^x \sigma_{i+1}^x + \sigma_i^y \sigma_{i+1}^y + \Delta (\sigma_i^z \sigma_{i+1}^z - 1)\right] {\cal P}_s,\\
\Delta &=& \frac{\epsilon_+ + \epsilon_-}{\sqrt{\epsilon_+ \epsilon_-}}.\nonumber
\eea
Apart from the projector this Hamiltonian coincides with the gapped ferromagnetic Heisenberg chain. However now, in distinction with \rf{e8}, the boundary condition is not periodic but twisted
\beq
\label{e11}
\sigma_{N+1}^{\pm} = (\frac{\epsilon_+}{\epsilon_-})^{\pm \frac{L}{2}} \sigma_1^{\pm}, \sigma_{N+1}^z = \sigma_1^z.
\eeq
Since $\frac{\epsilon_+}{\epsilon_-} \neq 1$ this boundary term has the same degree of importance than the whole system and we have a critical behavior induced by the surface, i. e., the mass gap vanishes in opposition to the standard periodic ferromagnetic XXZ chain.

\section{ The exact solution of the generalized asymmetric exclusion problem: The matrix product {\it ansatz}}

        The exact solution of the generalized asymmetric exclusion problem of last section was derived in~\cite{PRE} within the framework of the coordinate Bethe {\it ansatz}. In this section we are going to rederive this solution by imposing a matrix product {\it ansatz} for the eigenfunctions of the Hamiltonian \rf{e6}. As we shall see this derivation turns out to be more direct than the old one presented in~\cite{PRE}.

        Before considering the more general situation where any molecule may have a distinct size let us consider initially the simple case where all the molecules have the same size $s$ ($s=0,1,\ldots $).

        Since the diffusion process conserve particles, and the lattice is periodic, the total number of particles $n$ and the momentum $P$ are good quantum numbers. Consequently the Hilbert space associated to \rf{e6} can be separated into block disjoints sectors labelled by the values of $n$ ($n=0,1,\ldots $) and $P$ ($P=\frac{2\pi l}{L};l=0,1 2,\ldots ,L-1$). 

        Our {\it ansatz} asserts that any eigenfunction $|\Psi_{n,P} \rangle$ of \rf{e6} in the sector with $n$ particles and momentum $P$, will have its components given by the matrix product
\bea
\label{e12}
|\Psi_{n,P} \rangle &=& \sum_{\{ x_1, \ldots , x_n \}}^* f(x_1, \ldots , x_n)|x_1, \ldots , x_n\rangle, \\
f(x_1, \ldots , x_n) &=& \mbox{Tr}\left[E^{x_1-1}A^{(s)} E^{x_2-x_1-1}A^{(s)} \cdots E^{x_n-x_{n-1}-1}A^{(s)}E^{L-x_n} \Omega_P \right].\nonumber
\eea
The ket $|x_1, \ldots , x_n\rangle$ denotes the configuration where the particles are located at ($x_1, \ldots , x_n$) and the symbol ($*$) in the sum denotes the restriction to the sets satisfying the hard-core exclusion due to the size $s$ of the particles, i. e., 
\beq
\label{e13} 
x_{i+1} \geq x_i+s ,\;\;\; i = 1, \ldots , n-1 ,\;\;\; s \leq x_n - x_1 \leq L-s,
\eeq
where we have to remember that in the case where the particles have size $s=0$ we may have any number of particle in a given site.
Differently from the standard Bethe {\it ansatz} where $f(x_1, \ldots , x_n)$ is given by the combination of plane waves now it is given by the trace of a product of matrices. The matrices $E$ and $A^{(s)}$ are associated to the empty and occupied sites describing the configuration of the lattice. The superscript ($s$) is just to remember the size of the particle. The matrix $\Omega_P$ in \rf{e12} is introduced in order to ensure the momentum $P$ of the eigenfunction $|\Psi_{n,P} \rangle$. This is accomplished by imposing the commutation relation
\beq
\label{e14}
E\Omega_P = e^{-iP}\Omega_P E, \;\;\; A^{s} \Omega_P = e^{-iP}\Omega_P A^{s},
\eeq
since from \rf{e12} we must have for eigenfunctions of momentum $P$ the ratio of the amplitudes:
\beq
\label{e14b}
\frac{f(x_1, \ldots , x_n)}{f(x_1+m, \ldots , x_n+m)} = e^{-imP}, \;\;\; (m=1,2,\ldots ,L-1).
\eeq
The algebraic properties of $A^{(s)}$ and $E$ will be fixed by requiring that $|\Psi_{n,P} \rangle$, defining the {\it ansatz} \rf{e12}, satisfy the eigenvalue equation
\beq
\label{e15}
H_{\{ s_1=\cdots =s_n=s\} }|\Psi_{n,P} \rangle=\varepsilon_n |\Psi_{n,P} \rangle,
\eeq
where $H_{\{ s_1=\cdots =s_n=s\} }$ is given by \rf{e8}.

        Before considering the case where $n$ is general let us consider initially the cases where we have only $n=1$ or $n=2$ particles.

{\it {\bf  n = 1.}}
For one particle the eigenvalue equation \rf{e15} give us
\bea
\label{e15b}
\varepsilon_1 \mbox{Tr} \left( E^{x_1-1}A^{(s)} E^{L-x_1}\Omega_P\right) &=& - \epsilon_+ \mbox{Tr} \left( E^{x_1-2}A^{(s)}E^{L-x_1+1}\Omega_P \right) \nonumber \\
- \epsilon_- \mbox{Tr} \left( E^{x_1}A^{(s)} E^{L-x_1-1}\Omega_P\right) &+& (\epsilon_+ + \epsilon_-) \mbox{Tr} \left( E^{x_1-1}A^{(s)} E^{L-x_1}\Omega_P\right).
\eea
The cyclic property of the trace and the algebra \rf{e14} fix the values of the energies
\beq
\label{e16}
\varepsilon_1 = -(\epsilon_+ e^{-iP} + \epsilon_- e^{iP} -1),
\eeq
where $P = \frac{2\pi l}{L}$ ($l=0,1,\ldots ,L-1$), is the momentum of the state.

        An alternative way to solve \rf{e15b} that will be easier to generalize for arbitrary values of $n$ is obtained by the replacement
\beq
\label{e17} 
A^{(s)} = A_k^{(s)}E^{2-s},
\eeq
where $A_k$ is a spectral parameter dependent matrix with the following commutation relation with the matrix $E$:
\beq
\label{e18} 
EA_k^{(s)} =e^{ik} A_k^{(s)}E.
\eeq
Inserting \rf{e17} in \rf{e15b} and using \rf{e18} we obtain
\beq
\label{e19}
\varepsilon_1 = \varepsilon(k) = -(\epsilon_+ e^{-ik} + \epsilon_- e^{ik} -1),
\eeq
where we have used $\epsilon_+ + \epsilon_- = 1$.

        Comparing \rf{e19} with \rf{e16} we fix the spectral parameter $k$ as the momentum of the $1$-particle eigenfunction $|\Psi_{1,P} \rangle$, i. e., $k = P = \frac{2\pi l}{L}$ ($l=0,1, \ldots ,L-1$).

{\it {\bf  n =2.}} For two particles on the lattice the eigenvalue equation \rf{e15} gives for $|\Psi_{2,P} \rangle$ two types of relations depending on the relative location of the particles. The amplitudes corresponding to the configuration $|x_1,x_2 \rangle$ where $x_2>x_1+s$ will give the relation
\bea
\label{e20}
\varepsilon_2 \mbox{Tr} \left( E^{x_1-1}A^{(s)}E^{x_2-x_1-1}A^{(s)}E^{L-x_2}\Omega_P\right) &=& - \epsilon_+ \mbox{Tr} \left( E^{x_1-2}A^{(s)}E^{x_2-x_1}A^{(s)}E^{L-x_2}\Omega_P \right) \nonumber \\
- \epsilon_- \mbox{Tr} \left( E^{x_1}A^{(s)} E^{x_2-x_1-2}A^{(s)}E^{L-x_2}\Omega_P\right) &-&  \epsilon_+ \mbox{Tr} \left( E^{x_1-1}A^{(s)}E^{x_2-x_1-2}A^{(s)}E^{L-x_2+1}\Omega_P \right)\nonumber \\
- \epsilon_- \mbox{Tr} \left( E^{x_1-1}A^{(s)}E^{x_2-x_1}A^{(s)}E^{L-x_2-1}\Omega_P \right) &+& 2 \mbox{Tr} \left( E^{x_1-1}A^{(s)}E^{x_2-x_1-1}A^{(s)}E^{L-x_2}\Omega_P \right).
\eea
A possible and convenient way to solve this equation is by identifying the matrices $A^{(s)}$ as composed by two spectral parameter dependent new matrices $A_{k_1}^{(s)}$ and $A_{k_2}^{(s)}$, i. e.,
\beq
\label{e21} 
A^{(s)}=\sum_{i=1}^2 A_{k_i}^{(s)}E^{2-s},
\eeq
that satisfy, as in \rf{e18} the commutation relation
\beq
\label{e22}
E A_{k_j}^{(s)} = e^{ik_j} A_{k_j}^{(s)}E, \;\;\; (j=1,2).
\eeq
Inserting \rf{e21} in \rf{e20} and using \rf{e22} we obtain
\beq
\label{e23}
\varepsilon_2 = \varepsilon(k_1) + \varepsilon(k_2),
\eeq
where $\epsilon(k)$ is given in \rf{e19}.

The relation \rf{e14} give the commutation of these new matrices $A_{k_i}^{(s)}$ with $\Omega_P$, i. e.,
\beq
\label{e24}
A_{k_j}^{(s)} \Omega_P = e^{iP(1-s)} \Omega_P A_{k_j}^{(s)}, \;\;\; (j=1,2).
\eeq
Comparing the components of the configurations $|x_1,x_2\rangle$ and $|x_1+m,x_2+m\rangle$, and exploring the cyclic invariance of the trace we obtain 
\beq
\label{e25}
P = k_1 + k_2.
\eeq
Up to now the commutation relations of the matrices $A_{k_1}^{(s)}$ and $A_{k_2}^{(s)}$ among themselves as well  the spectral parameters, that in general may be complex, are unknown. The eigenvalue equation \rf{e15} when applied to the components of the configuration $|x_1,x_2\rangle$ where $x_2=x_1+s$ (``matching'' conditions) will give us the relation 
\bea
\label{e26}
\varepsilon_2 \mbox{Tr} \left( E^{x_1-1}A^{(s)}E^{s-1}A^{(s)}E^{L-x_1-s}\Omega_P\right) &=& - \epsilon_+ \mbox{Tr} \left( E^{x_1-2}A^{(s)}E^{s}A^{(s)}E^{L-x_1-s}\Omega_P \right) \nonumber \\
- \epsilon_- \mbox{Tr} \left( E^{x_1-1}A^{(s)} E^{s}A^{(s)}E^{L-x_1-s-1}\Omega_P\right)  &+& \mbox{Tr} \left( E^{x_1-1}A^{(s)}E^{s-1}A^{(s)}E^{L-x_1-s}\Omega_P \right).
\eea
Using \rf{e21} to express the $A^{(s)}$ matrix in terms of the spectral parameter matrices $A_{k_j}^{(s)}$ ($j=1,2$), and \rf{e19} for $\varepsilon_2$, the last expression gives 
\beq
\label{e27}
\sum_{j,l}^2 \left[ \epsilon_- - e^{-ik_j} + \epsilon_+ e^{-i(k_j +k_l)} \right] A_{k_j}^{(s)} A_{k_l}^{(s)}=0.
\eeq
This last relation imply that the matrices $\{A_{k_j}\}$ should obey the algebra 
\beq
\label{e28}
A_{k_j}^{(s)} A_{k_l}^{(s)} = S(k_j, k_l) A_{k_l}^{(s)} A_{k_j}^{(s)}, \;\;\; (l \neq j), \;\;\; \left( A_{k_j}^{(s)} \right)^2 = 0.
\eeq
where
\beq
\label{e29}
S(k_j, k_l) = - \frac{\epsilon_+ + \epsilon_-e^{i(k_j +k_l)} - e^{ik_j}}{\epsilon_+ + \epsilon_-e^{i(k_j +k_l)} - e^{ik_l}}.
\eeq
The complex spectral parameters $\{k_j\}$ are still free up to now. The cyclic property of the trace together with the algebraic relations \rf{e14}, \rf{e22}, \rf{e24} and \rf{e28} and the fact that any component should be uniquely related give us 
\bea
\label{e30}
\mbox{Tr} \left( A_{k_l}^{(s)}A_{k_j}^{(s)}E^{L-2s+2}\Omega_P\right) &=& e^{-i(L-2s+2)k_j} \mbox{Tr} \left( A_{k_l}^{(s)}E^{L-2s+2}A_{k_j}^{(s)}\Omega_P\right) \nonumber \\ 
&=& e^{-ik_j L} e^{i2k_j(s-1)}e^{-iP(s-1)} \mbox{Tr} \left( A_{k_j}^{(s)}A_{k_l}^{(s)}E^{L-2s+2}\Omega_P\right) \nonumber \\
&=& e^{-ik_j L} e^{i2k_j(s-1)}e^{-iP(s-1)} S(k_j, k_l) \mbox{Tr} \left( A_{k_l}^{(s)}A_{k_j}^{(s)}E^{L-2s+2}\Omega_P\right),
\eea
or equivalently, since $P=k_1+k_2$,
\beq
\label{e31}
e^{ik_j L} = S(k_j, k_l) \left( \frac{e^{ik_j}}{e^{ik_l}} \right)^{s-1},\;\;\; j=1,2\;\;\; (j \neq l).
\eeq
The energy and momentum are given by inserting the solution of \rf{e31} into \rf{e23} and \rf{e25}, respectively.

{\it {\bf General n.}} The above calculation can easily be extended to the case where $n>2$. The eigenvalue equation \rf{e15} when applied to the components of the eigenfunction corresponding to the configuration of $|\Psi_{n,P}\rangle$ where all the particles are at distances larger than the size $s$ of the particles, gives a generalization of \rf{e20}:
\bea
\label{e32}
\varepsilon_n \mbox{Tr} \left( \cdots E^{x_i-x_{i-1}-1}A^{(s)}E^{x_{i+1}-x_i-1}A^{(s)} \cdots A^{(s)} E^{L-x_n}\Omega_P\right) = \nonumber\\  
- \sum_{i=1}^n \{ \epsilon_+ \mbox{Tr} \left( \cdots E^{x_i-x_{i-1}-2}A^{(s)}E^{x_{i+1}-x_i}A^{(s)} \cdots A^{(s)}E^{L-x_n}\Omega_P \right) \nonumber \\
+ \epsilon_- \mbox{Tr} \left( \cdots E^{x_i-x_{i-1}-1}A^{(s)} E^{x_{i+1}-x_i-2}A^{(s)} \cdots A{(s)}E^{L-x_n+1}\Omega_P\right) \nonumber \\
-  \mbox{Tr} \left( \cdots E^{x_i-x_{i-1}-1}A^{(s)}E^{x_{i+i}-x_i-1}A^{(s)} \cdots A^{(s)}E^{L-x_n}\Omega_P \right) \}.
\eea
The solution is obtained by identifying the $A^{(s)}$ matrix as a combination of $n$ spectral parameter dependent $\{A_{k_j}^{(s)}\}$ matrices, namely,
\beq
\label{e33}
A^{(s)}=\sum_{j=1}^n A_{k_j}^{(s)}E^{2-s},
\eeq
with the commutation relations with the matrices $E$ and $\Omega_P$
\beq
\label{e34}
E A_{k_j}^{(s)} = e^{ik_j} A_{k_j}^{(s)}E, \;\;\; A_{k_j}^{(s)} \Omega_P = e^{iP(1-s)} \Omega_P A_{k_j}^{(s)} \;\;\; (j=1, \ldots ,n).
\eeq
Inserting \rf{e33} into \rf{e32} and using the relations \rf{e34}, together with the cyclic property of the trace we obtain 
\beq
\label{e34b}
\varepsilon_n = \sum_{j=1}^n \varepsilon(k_j),\;\;\; P = \sum_{j=1}^n k_j,
\eeq
for the energy and momentum of $|\Psi_{n,P}\rangle$, respctively.
The eigenvalue equation \rf{e15} applied to the configuration where a pair of particles located at $x_i$ and $x_{i+1}$ are at the closest position, i. e., $x_{i+1}=x_i+s$, will give relations that coincides with \rf{e25} and \rf{e29}, but now with $j=1,2,\ldots ,n$.
The configurations of $|\Psi_{n,P}\rangle$ corresponding to three or more particles at the ``matching'' distances will demand that the algebra satisfied by the matrices $\{A_{k_j}\}$ in \rf{e28} is associative. Equivalently this means that a given component, expressed in terms of a product of matrices $\{A_{k_j}\}$ and $E$, should be uniquely related to the other components. This is immediate for the present problem since the structure constants $S(k_j,k_l)$ of the algebra in \rf{e28} are constants with the property 
\beq
\label{e35}
S(k_j, k_l)S(k_l, k_j) = 1.
\eeq
As we are going to see in section $4$, this condition in general leads to the well known Yang-Baxter relations~\cite{yang2,baxter}.

Once the algebra is defined all the components of $|\Psi_{n,P}\rangle$ can be uniquely determined only if this algebra has a well defined trace, whose cyclic property will fix the $n$ complex spectral parameters $\{k_j\}$. An analog procedure as in \rf{e30} give us the constraints
\beq
\label{e36}
e^{ik_jL}=(-1)^n \prod_{l=1}^n \left( \frac{e^{ik_j}}{e^{ik_l}} \right)^{s-1} \frac{\epsilon_+ + \epsilon_-e^{i(k_j +k_l)} - e^{ik_j}}{\epsilon_+ + \epsilon_-e^{i(k_j +k_l)} - e^{ik_l}}.
\eeq
This equation coincides with the Bethe-{\it ansatz} equations derived in~\cite{PRE} through the coordinate Bethe {\it ansatz} method. Moreover, an arbitrary component $f(x_1, \ldots, x_n)$ of the wave function $|\Psi_{n,P}\rangle$ given in \rf{e12}, can be written as
\bea
\label{e37}
&&f(x_1, \ldots, x_n) = \nonumber \\
&&\sum_{i_1=1}^n \sum_{i_2=1}^n \cdots \sum_{i_n=1}^n \mbox{Tr} \left( E^{x_1-1}A_{k_{i_1}}^{(s)}E^{x_2-x_1+1-s}A_{k_{i_2}}^{(s)} \cdots E^{x_n-x_{n-1}+1-s}A_{k_{i_n}}^{(s)} E^{L-x_n+2-s}\Omega_P\right).
\eea
 Using the commutation relation \rf{e22} and the fact that $\left( A_{k_j}^{(s)} \right)^2 = 0$ ($j=1, \ldots ,n$) we can rewrite this last expression as
\bea
\label{e38}
&&f(x_1, \ldots, x_n) = \nonumber \\
&&\sum_{i_1, \ldots, i_n} e^{i[k_{i_1}(x_1-1)+k_{i_2}(x_2-1)+ \cdots +k_{i_n}(x_n-1)]} \mbox{Tr} \left( A_{k_{i_1}}^{(s)}E^{1-s}A_{k_{i_2}}^{(s)}E^{1-s} \cdots E^{1-s}A_{k_{i_n}}^{(s)} E^{L}\Omega_P\right).
\eea
Let us define the new matrices
\beq
\label{e39}
\tilde{A}_{k_j}^{(s)}=A_{k_j}^{(s)}E^{1-s}\;\;\; (j=1,\ldots ,n).
\eeq
It is simple to verify, from \rf{e34}, that they satisfy
\beq
\label{e40}
\tilde{A}_{k_j}^{(s)} \tilde{A}_{k_l}^{(s)} = \tilde{S}(k_j, k_l) \tilde{A}_{k_l}^{(s)} \tilde{A}_{k_j}^{(s)}, \;\;\; (j \neq l), \;\;\; \left( \tilde{A}_{k_j}^{(s)} \right)^2 = 0,
\eeq
where
\beq
\label{e41}
\tilde{S}(k_j, k_l)=S(k_j, k_l)\left( \frac{e^{ik_j}}{e^{ik_l}} \right)^{s-1}.
\eeq
Finally, in terms of these new matrices, and exploring the fact that $\left( \tilde{A}_{k_j}^{(s)} \right)^2 = 0$ we can write
\beq
\label{e42}
  f(x_1, \ldots, x_n) = \sum_{p_1, \ldots, p_n} e^{i[k_{p_1}(x_1-1)+k_{p_2}(x_2-1)+ \cdots +k_{p_n}(x_n-1)]} Tr \left( \tilde{A}_{k_{p_1}}^{(s)} \tilde{A}_{k_{p_2}}^{(s)} \cdots \tilde{A}_{k_{p_n}}^{(s)} E^{L}\Omega_P\right),
\eeq
where the sum is over the permutations ($p_1, p_2, \ldots,p_n$) of non repeated integers ($1,2, \ldots,n$). The result \rf{e42} show us that the amplitudes derived using the present matrix product {\it ansatz} is given by a combination of planes waves with complex wave number $\{k_j\}$, and reproduces the results obtained previously~\cite{PRE} through the standard coordinate Bethe {\it ansatz}. 

Let us return to the general case where we have $n$ molecules with arbitrary sizes $\{s_1,s_2, \ldots,s_n\}$ and whose related Hamiltonian is given by \rf{e6}. In this general case each particle is conserved separately, and since in the diffusion processes no interchange of particles are allowed, also the order $\{s_1,s_2, \ldots,s_n\}$ where the particles appear is a constant of motion, up to cyclic permutations. The eigenfunction corresponding to a given order $\{s_1,s_2, \ldots,s_n\}$ and momentum $P$ can be written as 
\beq
\label{e43}
|\Psi_{\{ s_1, \ldots, s_n\},P} \rangle = \sum_{\{c\}} \sum_{\{x\}} f^{s_{c_1}, \ldots, s_{c_n}}(x_1, \ldots, x_n) | x_1, \ldots, x_n \rangle,\eeq
where $f^{s_{c_1}, \ldots, s_{c_n}}(x_1, \ldots, x_n)$ is the component of a configuration where the particles of sizes $s_{c_1}, \ldots, s_{c_n}$ are located at positions $x_1, \ldots, x_n$ respectively. The summation $\{c\}$ extends over all cyclic permutations $\{c_1, \ldots, c_n\}$ of integers $\{1, \ldots, n\}$, and the summation $\{x\}$ extends, for a given distribution $\{s_{c_1}, \ldots, s_{c_n} \}$ of molecules, to increasing integers satisfying
\bea
\label{e44}
 x_{i+1}-x_i \geq s_{c_i},\;\;\; i=1, \ldots, n-1 \nonumber \\
 s_{c_1} \leq x_n - x_1 \leq N-s_{c_n}.
\eea

In order to formulate our matrix product {\it ansatz} we associate to the sites occupied by the particles of size $s_j$ ($j=1, \ldots, n$) a matrix $A^{(s_j)}$ and to the remaining $L-n$ sites we associate, as before, the matrix $E$. Our {\it ansatz} asserts, in a generalization of \rf{e12}, that the amplitudes of the eigenfunctions \rf{e43} are given by
\beq
\label{e45}
 f^{s_1, \ldots, s_n}(x_1, \ldots, x_n) = \mbox{Tr} \left( E^{x_1-1}A^{(s_1)}E^{x_2-x_1-1}A^{(s_2)} \cdots E^{x_n-x_{n-1}-1}A^{(s_n)}E^{L-x_n} \Omega_P \right),
\eeq
where in order to ensure the momentum $P$ of the eigenstate the matrices $\{ A^{\{ s \}}\}$ should satisfy
\beq
\label{e46}
E \Omega_P = e^{-iP} \Omega_P E,\;\;\; A^{(s)} \Omega_P = e^{-iP} \Omega_P A^{(s)}.
\eeq
Let us consider initially our {\it ansatz} \rf{e43}-\rf{e46} for $n=1$ and $n=2$ molecules.

{\it {\bf n=1}} For one particle on the chain we have the same {\it ansatz} as \rf{e12} and the energy given by \rf{e16}.

{\it {\bf n=2}} In this case if both particles have the same size $s_1 = s_2 = s$ we have the same situation considered previously in \rf{e20}-\rf{e31}. The eigenfunctions $|\Psi_{\{s,s\},P}\rangle$ will be given by \rf{e12} and the energy by \rf{e22} with ${k_i}$ fixed by \rf{e31}. 

        If the particles are distinct the situation is new. The eigenvalue equation when applied to the configurations where the two particles of sizes $s_1$ and $s_2$ are located at $x_1$ and $x_2\geq x_1 + s_1$, respectively, will give, {\it mutatis mutandis}, an expression similar to \rf{e30}. The corresponding situation are obtained by introducing the generalization of the spectral parameter matrices defined on \rf{e21}, i. e.,
\beq
\label{e47} 
A^{(s_j)}=\sum_{l=1}^2 A_{k_l}^{(s_j)}E^{2-s_j},\;\;\; (j=1,2),
\eeq
that satisfy, as \rf{e22}, the commutation relations
\beq
\label{e48}
E A_{k_l}^{(s_j)} = e^{ik_l} A_{k_l}^{(s_j)}E, \;\;\; (j,l=1,2).
\eeq
The energy $\varepsilon_{\{ s_1,s_2 \}}$ end momentum $P$ are related to the spectral parameters by \rf{e23} and \rf{e25}, respectively, and from \rf{e46} and \rf{e47}, we have the commutation relations, generalizing \rf{e24}
\beq
\label{e49}
A_{k_l}^{(s_j)} \Omega_P = e^{iP(1-s_j)} \Omega_P A_{k_l}^{(s_j)}, \;\;\; (j,l=1, \ldots ,n),
\eeq
with $n=2$.

        If the two particles are at the closest distance $x_2 = x_1 + s_1$ (``matching'' condition) the expression \rf{e26} should be replaced by
\bea
\varepsilon_{\{ s_1, s_2 \}} \mbox{Tr} \left( E^{x_1-1}A^{(s_1)}E^{s_1-1}A^{(s_2)}E^{L-x_1-s_1}\Omega_P\right) &=& - \epsilon_+ \mbox{Tr} \left( E^{x_1-2}A^{(s_1)}E^{s_1}A^{(s_2)}E^{L-x_1-s_1}\Omega_P \right) \nonumber \\
- \epsilon_- \mbox{Tr} \left( E^{x_1-1}A^{(s_1)} E^{s_1}A^{(s_2)}E^{L-x_1-s_1-1}\Omega_P\right)  &+& \mbox{Tr} \left( E^{x_1-1}A^{(s_1)}E^{s_1-1}A^{(s_2)}E^{L-x_1-s_1}\Omega_P \right). \nonumber 
\eea
Inserting the definition \rf{e47}, the expression \rf{e23} for $\varepsilon_{\{ s_1, s_2 \}}$ and using the algebraic relations \rf{e48} and \rf{e49} we obtain the commutation relations for the matrices $\{ A_{k_j}^{(s_l)} \}$
\beq
\label{e51}
A_{k_j}^{(s_l)} A_{k_m}^{(s_r)} = S(k_j, k_m) A_{k_m}^{(s_l)} A_{k_j}^{(s_r)}, \;\;\; (j \neq m ; l,r=1,2), \;\;\; A_{k_j}^{(s_1)} A_{k_j}^{(s_2)}= 0,
\eeq
where $S(k_j, k_m)$ is given by the same expression as \rf{e29}. It is interesting to notice that the structure constants $S(k_j, k_m)$ of the algebra in \rf{e51} are independent of the superscript of the matrices $A_{k_j}^{(s_l)}$, and consequently the algebra among the $\{ A_{k_j}^{(s_l)} \}$ is the same as that of \rf{e28} with respect to the interchange of spectral parameters. However the superscript of these matrices can not be neglected since in the commutations relations \rf{e51} they are not interchanged and also their commutation  with the $\Omega_P$ matrix is size dependent (see \rf{e49}).

        The spectral parameters $k_1$ and $k_2$ are fixed by the cyclic property of the trace, and we have, {\it mutatis mutandis}, a similar expression as \rf{e30}. Using the algebraic relations \rf{e48}, \rf{e49} and \rf{e51} we obtain
\bea
\label{e52}
&&\mbox{Tr} \left( A_{k_l}^{(s_1)}A_{k_j}^{(s_2)}E^{L-s_1-s_2+2}\Omega_P \right)= \nonumber \\ 
&&e^{-ik_jL}e^{ik_j(s_1+s_2-2)}e^{-iP(s_2-1)}S(k_j, k_l)\mbox{Tr} \left( A_{k_l}^{(s_1)}A_{k_j}^{(s_2)}E^{L-s_1-s_2+2}\Omega_P \right).
\eea
However differently from \rf{e30} the traces on the left and in the right hand side of the equation are not the same if $s_1 \neq s_2$. If we repeat once more the commutations that lead to \rf{e52} we obtain the same trace in both sides and consequently
\beq
\label{e53}
\left[ e^{-ik_jL}e^{ik_j(s_1+s_2-2)}S(k_j, k_l) \right]^2 e^{-iP(s_1+s_2-1)}=1.
\eeq
Since $P=k_1+k_2$, this last expression is equivalent to 
\beq
\label{e54}
e^{ik_jL} = e^{i \frac{2 \pi}{2}m}\left( \frac{e^{ik_j}}{e^{ik_l}} \right)^{\bar{s}-1}S(k_j, k_l), \;\;\; m = 0,1; \;\;\; j \neq l =1,2 ; \;\;\; s_1 \neq s_2,
\eeq
and 
\beq
\label{e55}
\bar{s} = \frac{s_1+s_2}{2}
\eeq
is the average size of the two molecules. The expression \rf{e54} generalizes the expression \rf{e31} obtained for particles of equal sizes. We note however that since $m = 0,1$ in \rf{e54} we have two times more solutions than the corresponding one \rf{e31} for particles of equal sizes. This indeed should be the case since particles of distinct sizes are distinguishable, and consequently the number of eigenfunction is doubled when compared with the indistinguishable case ($s_1 = s_2$).

{\it {\bf General n.}} In this case we have a general distribution of particles with sizes $\{ s_1, s_2, \ldots , s_n \}$ and the corresponding eigenfunctions are given by \rf{e43} and \rf{e45}. The eigenvalue equation when applied to the components $|x_1, \ldots , x_n \rangle$ where all the  particles are not at the closest distance, gives an equation similar to \rf{e32}, {\it mutatis mutandis}, whose solution is given by the generalization of \rf{e47}, \rf{e49} and \rf{e48}:
\beq
\label{e56}
A^{(s_j)}=\sum_{l=1}^n A_{k_l}^{(s_j)}E^{2-s_j},\;\;\; A_{k_l}^{(s_j)} \Omega_P = e^{iP(1-s_j)} \Omega_P A_{k_l}^{(s_j)}, \;\;\;  E A_{k_l}^{(s_j)} = e^{ik_l} A_{k_l}^{(s_j)}E, \;\;\; (j=1,2),
\eeq
producing the energy and momentum given by
\beq
\label{e57}
\varepsilon_n = \sum_{j=1}^n \left( \epsilon_+ e^{-ik_j} + \epsilon_- e^{ik_j} -1 \right), \;\;\; P = \sum_{j=1}^n k_j,
\eeq
respectively. The eigenvalue equation applied to the components where a pair of particles ($x_i, x_{i+1}$) are located at the closest distance, $ x_{i+1} = x_i+s_i$ will give a generalization of \rf{e51}
\beq
\label{e58}
A_{k_j}^{(s_t)} A_{k_l}^{(s_u)} = S(k_j, k_l) A_{k_l}^{(s_t)} A_{k_j}^{(s_u)}, \;\;\; (j \neq l), \;\;\; A_{k_j}^{(s_t)} A_{k_j}^{(s_u)}= 0,  \;\;\; (j,l,t,u=1, \ldots ,n). 
\eeq
The cyclic property of the trace in \rf{e45} will give, by using \rf{e56} and \rf{e58} a generalization of \rf{e52}, namely, for each $k_j$,
\bea
\label{e59}
&&\mbox{Tr} \left(A_{k_1}^{(s_1)} A_{k_2}^{(s_2)} \cdots A_{k_{j-1}}^{(s_{j-1})}  A_{k_j}^{(s_j)} \cdots  A_{k_n}^{(s_n)} E^{L-\sum_{i=1}^n (s_i-1)}\Omega_P \right)= e^{-ik_jL}e^{ik_j \sum_{i=1}^n (s_i-1)}e^{-iP(s_j-1)}\times \nonumber \\ 
&& \left( \prod_{l=1}^n S(k_j, k_l) \right) \mbox{Tr} \left(A_{k_1}^{(s_n)} A_{k_2}^{(s_1)} \cdots A_{k_{j}}^{(s_{j-1})}  A_{k_{j+1}}^{(s_j)} \cdots  A_{k_n}^{(s_{n-1})} E^{L-\sum_{i=1}^n (s_i-1)}\Omega_P \right).
\eea
Similarly as happened in \rf{e52} the traces in both sides of the last equation are not the same because $\{ s_1, s_2, \ldots , s_n \} \neq \{ s_n, s_1, \ldots , s_{n-1} \}$. But we can redue the above commutations by $r$ times until we reach the same distribution of sizes, where $r$ is the minimum number of cyclic rotations of  $\{ s_1, s_2, \ldots , s_n \}$ where the configuration repeats the initial one. In this case we obtain
\beq
\label{e60}
\left[ e^{-ik_jL}e^{ik_j \sum_{i=1}^n (s_i-1)} \prod_{l=1}^n S(k_j, k_l) \right]^r e^{-iP \frac{r}{n} \sum_{i=1}^n (s_i-1)} = 1.
\eeq
Since $P = \sum_{i=1}^n k_j$, we can rewrite this last expression as 
\beq
\label{e61}
e^{ik_jL} = e^{i \frac{2 \pi}{r} m}  \prod_{l=1}^n S(k_j, k_l) \left( \frac{e^{ik_j}}{e^{ik_l}} \right)^{\bar{s}-1}, \;\;\; m=0,1,\ldots, r-1, \;\;\; j,l=1, \ldots , n,
\eeq 
where as in \rf{e55}
\beq
\label{e62}
\bar{s} = \frac{1}{n} \sum_{i=1}^n s_i
\eeq
is the average size of the molecules in the distribution $\{ s_1, s_2, \ldots , s_n \}$. This equation gives us a number of solutions of order $r$ times larger than the corresponding number in the case where all the particles have the same size. 

        The equation \rf{e61} that fix the spectral parameters of the matrices coincides with the Bethe -{ \it ansatz} equations derived in~\cite{PRE}. Similarly as we did in \rf{e37} - \rf{e42} we can show that indeed the eigenfunctions we obtained by using our matrix product { \it ansatz} coincides with the ones derived in the framework of the coordinate Bethe { \it ansatz}.

\section{ The asymmetric diffusion model with $N$ classes of particles with hierarchical order}

        The extension of the simple exclusion problem to the case where we have $N$ distinct classes of particles ($c = 1,2, \ldots ,N$) diffusing asymmetrically is not exact integrable in general. However the integrability of the problem can be preserved if the diffusive transitions of the several species happen in an hierarchical order. This problem was considered originally in the case $N=2$ as a model to describe shocks~\cite{bold}-\cite{ferra2} in nonequilibrium. The stationary properties of the $N=2$~\cite{derr1} and $N=3$~\cite{mallick} models can also be studied through a matrix product { \it ansatz}. In~\cite{BJP2} a generalization of this problem was considered in which the particles in each of the $N$ classes ($c = 1,2, \ldots ,N$) may have distinct sizes ($s_1, \ldots ,s_N$), respectively. The solution of this generalized problem was obtained through the coordinate Bethe { \it ansatz}~\cite{BJP2}. In the next section we are going to show that the solution of this problem, similarly as we did in the last section, can also be obtained through an appropriate matrix product { \it ansatz}.

        In this generalized problem we consider the particles in each class $c$ as composed by $s_c$ monomers, thus occupying $s_c$ sites on the lattice ($c=1,2, \ldots$). We consider as the position of the molecule the coordinate of its leftmost monomer. The excluded volume of a particle in class $c$ is given by its size $s_c$ ($c = 1,2, \ldots ,N$) in units of lattice spacing. The configurations of the molecules in the lattice is described by defining at each lattice site $i$ a variable $\beta_i$ ($i=1,2, \ldots ,L$), taking the values $\beta_i = 0,1, \ldots ,N$. The values $\beta_i = 1, \ldots ,N$ represent sites occupied by molecules of class $c = 1,2, \ldots ,N$, respectively. The sites attached with the value $\beta = 0$ are the vacant sites or those excluded due to the size of the particles. As an example $\{ \beta \} = \{ 1,0,2,0,2,0 \}$ may represent the configuration where in a $L = 6$ sites we have a particle of class $1$ and size $s_1 = 2$ at the site $1$ and two particles of class $2$ and size $s_2 = 1$ located at the sites $3$ and $5$. The allowed configurations are given, in general, by the set $\{ \beta_i \}$ ($i = 1,2, \ldots ,L$), where for each pair $(\beta_1 , \beta_j) \neq 0$ with $j>i$ we have $j-i > s_{\beta_i}$. The allowed stochastic process in the problem are just given by the exchange of particles or the asymmetric diffusion if the constraint due to the size of particles is satisfied. The possible motions of a given molecule are diffusion to the right
\beq
\label{e63}
\beta_i\;\;\emptyset_{i+1} \rightarrow \emptyset_i\;\;\beta_{i+1} \;\;\;\;\;\; (\mbox{rate} \;\; \Gamma_R),
\eeq
diffusion to the left
\beq
\label{e64}
\emptyset_i\;\;\beta_{i+1} \rightarrow \beta_i\;\;\emptyset_{i+1} \;\;\;\;\;\; (\mbox{rate} \;\; \Gamma_L),
\eeq
and interchange of particles
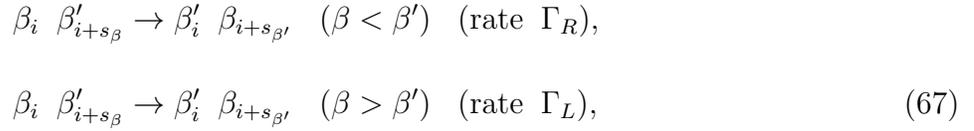
\bea
\label{e65}
\beta_i\;\;\beta'_{i+s_{\beta}} &\rightarrow& \beta'_i\;\;\beta_{i+s_{\beta'}}\;\;\; (\beta < \beta') \;\;\; (\mbox{rate} \;\; \Gamma_R), \nonumber \\
\beta_i\;\;\beta'_{i+s_{\beta}} &\rightarrow& \beta'_i\;\;\beta_{i+s_{\beta'}}\;\;\; (\beta > \beta') \;\;\; (\mbox{rate} \;\; \Gamma_L),
\eea
with $\beta , \beta' = 1,2, \ldots ,N$. As we see from \rf{e65}, particles of a class $c$ interchange positions with those in classes $c' > c$ with the same rate as they interchange positions with the vacant sites (diffusion). However the net effect of these motions is distinct from the diffusion processes, since by interchanging positions, distinctly from the diffusion process, the particles moves by $s_{c'}$ lattice units, accelerating its diffusion if $s_{c'}>1$. The identification of the master equation as an Schr\"odinger equation as in \rf{e5} will give us the Hamiltonian~\cite{BJP2}
\bea
\label{e66}
H &=& D \sum_{j=1}^L H_j \nonumber \\
H_j &=& -{\cal P} \{ \sum_{\alpha=1}^N \left[ \epsilon_+ (E_j^{0, \alpha}E_{j+1}^{\alpha, 0}-E_j^{\alpha, \alpha}E_{j+1}^{0, 0}) + \epsilon_- (E_j^{\alpha, 0}E_{j+1}^{0, \alpha}-E_j^{0, 0}E_{j+1}^{\alpha, \alpha}) \right] \nonumber \\
&&+ \sum_{\alpha=1}^N \sum_{\beta=1}^N \epsilon_{\alpha, \beta} (E_j^{\beta, \alpha}E_{j+s_{\beta}}^{\alpha, 0}E_{j+s_{\alpha}}^{0, \beta}-E_j^{\alpha, \alpha}E_{j+s_{\beta}}^{0, 0}E_{j+s_{\alpha}}^{\beta, \beta}) \} {\cal P},
\eea
with
\beq
\label{e67}
D = \Gamma_R + \Gamma_L, \;\;\; \epsilon_+ = \frac{\Gamma_R}{\Gamma_R +\Gamma_L}, \;\;\; \epsilon_- = \frac{\Gamma_L}{\Gamma_R +\Gamma_L} \;\;\; (\epsilon_+ + \epsilon_- = 1),
\eeq
\beq
\label{e68}
$$\epsilon_{\alpha \beta} = \left\{ \matrix{ \epsilon_+ & \alpha < \beta \cr
                                                   0          & \alpha = \beta \cr
                                                   \epsilon_- & \alpha > \beta    }
                            \right.$$
\eeq
and periodic boundary conditions. The matrices $E^{\alpha, \beta}$ are $(N+1) \times (N+1)$-dimensional matrices with a single nonzero element $(E^{\alpha, \beta})_{ij} = \delta_{\alpha i} \delta_{\beta j}$ ($\alpha , \beta , i , j = 0, \ldots ,N $). The projector ${\cal P}$ in \rf{e66} projects out the configurations associated to the vectors $| \{ \beta \} \rangle$ representing molecules at forbidden positions due to their finite size. Mathematically this condition means that for all $i , j$ with $\beta_i , \beta_j \neq 0$ we should have $|i-j| \geq s_{\beta_i}$ ($j > i$). The constant $D$ in \rf{e66} fixes the time scale in the problem and we chose $D = 1$. The Hamiltonian \rf{e66} corresponding to the particular case where all the molecules have unit size is related to the spin-$\frac{N}{2}$ SU($N+1$) anisotropic Sutherland chain~\cite{sutherland,tsveli} or SU($N+1$) Perk-Schultz model~\cite{perkshultz} with twisted boundary conditions~\cite{alcrit1}.

        At the end of the next section we are going to present the solution of a even further generalized model whose solution were not derived in~\cite{BJP2}. The solution of this model is quite complicated through the standard coordinate Bethe { \it ansatz}. As we shall see, nowever its derivation is not difficult through our matrix product { \it ansatz}. In this generalization instead of having all the particles in a given class $c$ with fixed size $s_c$, each individual particle may have an arbitrary size. In this case the configurations on the lattice are given by $\{ \vec{\beta} \} = \{ \vec{\beta}_1 , \vec{\beta}_2, \ldots , \vec{\beta}_L  \}$ where $\vec{\beta}_i = (c , s)$ mean that the lattice site $i$ ($i = 1,2, \ldots ,L$) is occupied by a particle of class $c$ ($c = 1,2, \ldots ,N$) having size $s$ ($s = 1,2, \ldots $). The Hamiltonian related to this stochastic problem is given by a generalization of \rf{e66}, namely
\bea
\label{e69}
H &=& D \sum_{j=1}^L H_j \nonumber \\
H_j &=& -{\cal P} \left\{ \sum_{\vec{\beta}} \left[ \epsilon_+ (E_j^{\vec{0}, \vec{\beta}}E_{j+1}^{\vec{\beta}, \vec{0}}-E_j^{\vec{\beta}, \vec{\beta}}E_{j+1}^{\vec{0}, \vec{0}}) + \epsilon_- (E_j^{\vec{\beta}, \vec{0}}E_{j+1}^{\vec{0}, \vec{\beta}}-E_j^{\vec{0}, \vec{0}}E_{j+1}^{\vec{\beta}, \vec{\beta}}) \right] \right. \nonumber \\
&&+ \left. \sum_{\vec{\beta}=(c , s)} \sum_{\vec{\beta}'=(c' , s')} \epsilon_{c c'} (E_j^{\vec{\beta}', \vec{\beta}}E_{j+s'}^{\vec{\beta}, \vec{0}}E_{j+s}^{\vec{0}, \vec{\beta}'}-E_j^{\vec{\beta}, \vec{\beta}}E_{j+s'}^{\vec{0}, \vec{0}}E_{j+s}^{\vec{\beta}', \vec{\beta}' }) \right\} {\cal P},
\eea
with $\epsilon_+$, $\epsilon_-$ and $\epsilon_{c c'}$ ($c, c' =1,2, \ldots ,N$) given as in \rf{e67} and \rf{e68}.

\section{ A matrix product {\it ansatz}  for the generalized diffusion problem with $N$ classes of particles with hierarchical order }

        The exact solution of the asymmetric diffusion problem with $N$ classes of particles, whose related Hamiltonian is given by \rf{e67} was obtained in~\cite{BJP2} through the coordinate Bethe {\it ansatz}. In this section we are going to reobtain this solution by an appropriate matrix product {\it ansatz}. Moreover our solution enables the extension to the more general problem discussed in the last section and whose Hamiltonian was introduced in \rf{e69}.
 
        Let us consider initially the simple case where all the particles in a given class ($c=1, \ldots ,N$) have a fixed size ($s_c = 1,2, \ldots$). Due to the conservation of particles in the diffusion and interchange processes the total number of particles in each class is conserved separately and we can split the associated Hilbert space into block disjoint eigensectors labeled by the numbers $n_1, n_2, \ldots ,n_N$ ($n_i =0,1, \ldots$) of particles on the classes $i$ ($i=1,2, \ldots ,N$). We want to obtain the eigenfunctions $|n_1, \ldots ,n_N \rangle$ of the eigenvalue equation
\beq
\label{e70}
H |n_1, \ldots , n_N \rangle = \varepsilon_n |n_1, \ldots , n_N \rangle,
\eeq
where
\beq
\label{e71}
|n_1, \ldots , n_N \rangle = \sum_{\{ c \}} \sum_{\{ x \}}^* f(x_1, c_1; \ldots ; x_n , c_n)|x_1, c_1; \ldots ; x_n , c_n \rangle.
\eeq
The ket $|x_1, c_1; \ldots ; x_n , c_n \rangle$ means the configuration where particles of class $c_i$ ($c_i = 1, \ldots ,N$) is located at position $x_i$ ($x_i = 1, \ldots ,L$) and the total number of particles is $n = n_1 + \cdots + n_N$. The summation $\{ c \}=\{ c_1, \ldots ,c_n \}$ extends over all the permutations of $n$ integer numbers $\{ 1,2, \ldots ,N \}$ in which $n_i$ terms have the value $i$ ($i = 1, \ldots , N$), while the summation $\{ x \}=\{ x_1, \ldots ,x_n \}$ runs, for each permutation $\{ c \}$, into the set of the nondecreasing integers satisfying
\bea
\label{e72}
x_{i+1} &\geq& x_i + s_{c_i}, \;\;\; i=1, \ldots ,n-1 \nonumber \\
s_{c_1} &\leq& x_n - x_1 \leq L - s_{c_n}.
\eea

        The matrix product { \it ansatz} we propose asserts that an arbitrary eigenfunction $|n_1, \ldots , n_N \rangle$ with momentum $P$, will have the amplitudes in \rf{e71} given in terms of traces of the matrix product 
\beq
\label{e73}
f(x_1, c_1; \ldots ; x_n , c_n) = \mbox{Tr} \left[ E^{x_1-1}Y^{(c_1)}E^{x_2-x_1-1}Y^{(c_2)} \cdots E^{x_n-x_{n-1}-1}Y^{(c_n)}E^{L-x_n} \Omega_P \right].
\eeq
The matrices $Y^{(c)}$ ($c=1, \ldots ,N$), $E$ and $\Omega_P$ will obey algebraic relations that ensure the validity of the eigenvalue equation \rf{e70}. The momentum $P$ of the state, analogously as in Sec. $3$ is fixed by requiring the relation 
\beq
\label{e74}
E \Omega_P = e^{-iP} \Omega_P E, \;\;\; Y^{(c)} \Omega_P = e^{-iP} \Omega_P Y^{(c)}, \;\;\; c = 1, \ldots ,N.
\eeq

        Let us consider the simplest cases of $n=1$ and $n=2$ particles before consider the case where $n$ is general.

{\it {\bf n=1}} In this case the problem is the same as that of Sec. $3$ and we obtain the energies given by \rf{e16}.

{\it {\bf n=2}} For two particles of classes $c_1$ and $c_2$ ($c_1, c_2 = 1, \ldots ,N$) on the lattice we have two distinct types of relations depending if the amplitudes are related or not to the configurations where two particles are at the closest distance $x_2=x_1+s_{c_1}$. The eigenvalue equation when applied to the components where the particles of class $c_1$ and $c_2$ are at positions ($x_2,x_1$), such that $x_2>x_1+s_{c_1}$, give us the relation
\bea
\label{e75}
&&\varepsilon_2  \mbox{Tr} \left[ E^{x_1-1}Y^{(c_1)}E^{x_2-x_1-1}Y^{(c_2)}E^{L-x_2} \Omega_P \right] = -\epsilon_+ \mbox{Tr} \left[ E^{x_1-2}Y^{(c_1)}E^{x_2-x_1}Y^{(c_2)}E^{L-x_2} \Omega_P \right] \nonumber \\
&& - \epsilon_-  \mbox{Tr} \left[ E^{x_1}Y^{(c_1)}E^{x_2-x_1-2}Y^{(c_2)}E^{L-x_2} \Omega_P \right] - \epsilon_+ \mbox{Tr} \left[ E^{x_1-1}Y^{(c_1)}E^{x_2-x_1-2}Y^{(c_2)}E^{L-x_2+1} \Omega_P \right] \nonumber \\
&& -\epsilon_- \mbox{Tr} \left[ E^{x_1-1}Y^{(c_1)}E^{x_2-x_1}Y^{(c_2)}E^{L-x_2-1} \Omega_P \right] + 2 \mbox{Tr} \left[ E^{x_1-1}Y^{(c_1)}E^{x_2-x_1-1}Y^{(c_2)}E^{L-x_2} \Omega_P \right].
\eea
A solution of this equation is obtained by identifying the matrices $Y^{(c)}$ as composed by two spectral parameter dependent new matrices $Y_{k_1}^{(c)}$ and $Y_{k_2}^{(c)}$, i. e. , 
\beq
\label{e76}
Y^{(c)} = \sum_{i=1}^2 Y_{k_i}^{(c)} E^{2- s_c}
\eeq
that, as in \rf{e22}, satisfy the commutation relation 
\beq
\label{e77}
E Y_{k_j}^{(c)} = e^{ik_j} Y_{k_j}^{(c)} E.
\eeq
In terms of the unknown spectral parameters $k_j$ ($j=1,2$) the energy and momentum are given by
\beq
\label{e78}
\varepsilon_2 = \varepsilon(k_1) + \varepsilon(k_2), \;\;\; P = k_1 + k_2,
\eeq
where $\varepsilon(k) = -(\epsilon_+ e^{-ik} + \epsilon_- e^{ik} - 1)$. As a consequence of \rf{e74} and \rf{e76} we also have
\beq
\label{e79}
Y_{k_j}^{(c)} \Omega_P = e^{iP(1-s_c)} \Omega_P Y_{k_j}^{(c)} \;\;\; (j=1,2;\;\; c=1, \ldots ,N). 
\eeq
The eigenvalue equation \rf{e70} when applied to the components of \rf{e71} where the two particles are at the closest distance, i. e., $x_2=x_1+s_{c_1}$ give us
\bea
\label{e80}
&&\varepsilon_2  \mbox{Tr} \left[ E^{x_1-1}Y^{(c_1)}E^{s_{c_1}-1}Y^{(c_2)}E^{L-x_1-s_{c_1}} \Omega_P \right] = -\epsilon_+ \mbox{Tr} \left[ E^{x_1-2}Y^{(c_1)}E^{s_{c_1}}Y^{(c_2)}E^{L-x_1-s_{c_1}} \Omega_P \right] \nonumber \\
 &-&\epsilon_-  \mbox{Tr} \left[ E^{x_1-1}Y^{(c_1)}E^{s_{c_1}}Y^{(c_2)}E^{L-x_1-s_{c_1}-1} \Omega_P \right] - \epsilon_{c_2 c_1} \mbox{Tr} \left[ E^{x_1-1}Y^{(c_2)}E^{s_{c_2}-1}Y^{(c_1)}E^{L-x_1-s_{c_2}} \Omega_P \right] \nonumber \\
&&+ 2 (1+\epsilon_{c_1 c_2}) \mbox{Tr} \left[ E^{x_1-1}Y^{(c_1)}E^{s_{c_1}-1}Y^{(c_2)}E^{L-x_1-s_{c_1}} \Omega_P \right].
\eea

        Substituting \rf{e76} and \rf{e78} in this last expression and using \rf{e77} we obtain
\bea
\label{e81}
&&\sum_{l,m} \left\{ \left[ -(\epsilon_+ e^{-i(k_l+k_m)} + \epsilon_-) + e^{-ik_l} (1 - \epsilon_{c_1,c_2}) \right] \mbox{Tr} \left[ E^{x_1}Y_{k_l}^{(c_1)}Y_{k_m}^{(c_2)}E^{L-x_1-s_{c_1}-s_{c_2}+2} \Omega_P \right] \right. \nonumber \\
&&+ \left. \epsilon_{c_2,c_1} e^{-ik_l} \mbox{Tr} \left[ E^{x_1}Y_{k_l}^{(c_2)}Y_{k_m}^{(c_1)}E^{L-x_1-s_{c_1}-s_{c_2}+2} \Omega_P \right] \right\} = 0.
\eea
This last equation is satisfied by imposing the following commutation relations among the operators $\{ Y_{k}^{(c)} \}$
\beq
\label{e82}
\sum_l \sum_m \left\{ \left[ {\cal D}_{l,m} +  e^{ik_m} (1 - \epsilon_{c_1,c_2}) \right] Y_{k_l}^{(c_1)}Y_{k_m}^{(c_2)} + \epsilon_{c_2,c_1} e^{ik_m} Y_{k_l}^{(c_2)}Y_{k_m}^{(c_1)} \right\} = 0,
\eeq
where
\beq
\label{e83}
{\cal D}_{l,m} = - ( \epsilon_+ + \epsilon_- e^{i(k_l+k_m)} ).
\eeq
It is interesting to consider separately the cases where the two particles belong to the same class $c_1 = c_2$ from the case where $c_1 \neq c_2$. If $c_1=c_2=c$ ($c=1, \ldots ,N$), since $\epsilon_{c,c}=0$ and ${\cal D}_{l,m}+e^{ik_m}\neq {\cal D}_{m,l}+e^{ik_l}$ for $l \neq m$ we obtain from \rf{e82} and \rf{e83}
\beq
\label{e84}
Y_{k_l}^{(c)}Y_{k_m}^{(c)} = S_{c,c}^{c,c}(k_l,k_m) Y_{k_m}^{(c)}Y_{k_l}^{(c)} \;\;\; (l \neq m), \;\;\; \left( Y_{k_l}^{(c)} \right)^2 = 0,
\eeq
where
\beq
\label{e85}
S_{c,c}^{c,c}(k_l,k_m) = - \frac{\epsilon_+ + \epsilon_- e^{i(k_l+k_m)} - e^{ik_l}}{\epsilon_+ + \epsilon_- e^{i(k_l+k_m)} - e^{ik_m}},
\eeq
and ($l,m=1,2;\;\; c=1, \ldots ,N$). The relation \rf{e82} in the cases where $c_1 \neq c_2$ give us the equations in matrix form
\beq
\label{e86}
\sum_{l,m=1}^2 
$$\left[ \matrix{  {\cal D}_{l,m} + \epsilon_{c_2,c_1} e^{ik_m} & \epsilon_{c_2,c_1} e^{ik_m} \cr
                        \epsilon_{c_1,c_2} e^{ik_m} & {\cal D}_{l,m} + \epsilon_{c_1,c_2} e^{ik_m}   }       \right] 
\left[ \matrix{ Y_{k_l}^{(c_1)}Y_{k_m}^{(c_2)} \cr
                Y_{k_l}^{(c_2)}Y_{k_m}^{(c_1)}     }
\right] = 0.$$
\eeq
Similarly as in~\cite{BJP2} the above equation can be rearranged straightforwardly by giving us the algebraic relations
\bea
\label{e87}
Y_{k_l}^{(c_1)}Y_{k_m}^{(c_2)} &=& \sum_{c'_1, c'_2 = 1}^N S_{c'_1,c'_2}^{c_1,c_2}(k_l,k_m) Y_{k_m}^{(c'_2)}Y_{k_l}^{(c'_1)} \;\;\; (k_l \neq k_m), \nonumber \\
Y_{k_l}^{(c_1)}Y_{k_l}^{(c_2)} &=& 0,
\eea
where ($l,m=1,2$), $c_1$, $c_2=1, \ldots ,N$ and the ``structure constants'' of the algebra are the components of an $S$-matrix whose non-zero components are given by \rf{e85} and
\bea
\label{e88}
S_{c_2,c_1}^{c_1,c_2} (k_1,k_2) &=&  [1 - \epsilon_{c_1,c_2} \Phi (k_1,k_2)] S_{c_1,c_1}^{c_1,c_1} (k_1,k_2) \;\;\; ( c_1,c_2 = 1, \ldots ,N ) \nonumber \\
S_{c_1,c_2}^{c_1,c_2} (k_1,k_2) &=& \epsilon_{c_2,c_1} \Phi (k_1,k_2) S_{c_1,c_1}^{c_1,c_1} (k_1,k_2) \;\;\; ( c_1,c_2 = 1, \ldots ,N; \;\; c_1 \neq c_2 ),
\eea
where
\beq
\label{e89}
\Phi (k_1,k_2) = \frac{e^{ik_1} - e^{ik_2}}{e^{ik_1} - \epsilon_+ - \epsilon_- e^{(ik_1+k_2)}}.
\eeq
The complex parameters ($k_1,k_2$), that are free up to now, are going to be fixed by the cyclic property of the trace in \rf{e73} and the algebraic relations \rf{e74}, \rf{e76}, \rf{e77} and \rf{e79}.

        Instead of solving for the spectral parameters for the $n=2$ let us consider the case of general $n$.

    

{\it {\bf General n.}} In this case the eigenvalue equation \rf{e70}, when applied to the components of the eigenfunction corresponding to the configuration where all the particles are at distances larger than the closest distance, give us a generalization of \rf{e75} that is promptly solved by identifying, as in \rf{e76}, the matrices $Y^{(c)}$ as combinations of $n$ spectral parameter matrices, i. e.,
\beq
\label{e90}
Y^{(c)} = \sum_{i=1}^n Y_{k_i}^{(c)} E^{2-s_c},
\eeq
satisfying the following algebraic relations with the matrices $E$,
\beq
\label{e91}
E Y_{k_j}^{(c)} = e^{ik_j} Y_{k_j}^{(c)} E \;\;\; (j=1, \ldots ,n; \;\; c=1,2, \ldots ,N),
\eeq
and from \rf{e74}
\beq
\label{e92}
Y_{k_j}^{(c)} \Omega_P = e^{iP(1-s_c)} \Omega_P Y_{k_j}^{(c)} E \;\;\; (j=1, \ldots ,n; \;\; c=1,2, \ldots ,N).
\eeq
The energy and momentum in terms of the spectral parameter $\{ k_j \}$ are given by the generalizations of \rf{e78}, namely,
\beq
\label{e93}
\varepsilon_n = \sum_{j=1}^{n} \varepsilon(k_j), \;\;\; P = \sum_{j=1}^{n} k_j.
\eeq
The components of the eigenfunctions corresponding to the configurations where a pair of particles of classes $c_1$ and $c_2$ are located at the closest positions $x_i$ and $x_{i+1} = x_i+s_{c_1}$, will give relations that reproduces \rf{e87} - \rf{e89}.

        Since in the general case we have the product of $n$ operators $\{ Y_{k_j}^{(c)} \}$, the algebraic relations \rf{e74}, \rf{e77}, \rf{e79} and \rf{e87} should provide a unique relation among these products. For example the product $\cdots Y_{k_1}^{(\alpha)}Y_{k_2}^{(\beta)}Y_{k_3}^{(\gamma)} \cdots$ can be related to the product $\cdots Y_{k_3}^{(\gamma)}Y_{k_2}^{(\beta)}Y_{k_1}^{(\alpha)} \cdots$ by two distinct ways. Either by performing the commutations $\alpha \beta \gamma \rightarrow \beta \alpha \gamma \rightarrow \beta \gamma \alpha \rightarrow \gamma \beta \alpha$ or by $\alpha \beta \gamma \rightarrow \alpha \gamma \beta \rightarrow \gamma \alpha \beta \rightarrow \gamma \beta \alpha$. Consequently we should have
\bea
\label{e94}
\sum_{\gamma, \gamma', \gamma''=1}^N && S_{\gamma, \gamma'}^{\alpha, \alpha'} (k_1,k_2) S_{\beta, \gamma''}^{\gamma, \alpha''} (k_1,k_3) S_{\beta',\beta''}^{\gamma', \gamma''} (k_2,k_3) = \nonumber \\
&&\sum_{\gamma, \gamma', \gamma''=1}^N S_{\gamma', \gamma''}^{\alpha', \alpha''} (k_2,k_3) S_{\gamma, \beta''}^{\alpha, \gamma''} (k_1,k_3) S_{\beta,\beta'}^{\gamma, \gamma'} (k_1,k_2),
\eea
for $\alpha, \alpha', \alpha'', \beta, \beta', \beta''=1, \ldots ,N$. This last relation is just the Yang-Baxter relations~\cite{yang2,baxter} of the $S$-matrix defined in \rf{e85} and \rf{e88}. Actually the condition \rf{e94} is enough to ensure that any matrix product is uniquely related and it implies the associativity of the algebra of the operators $\{ Y_{k_j}^{(c)} \}$. We can verify that the Yang-Baxter relation \rf{e94}, with $S$ given by \rf{e85} and \rf{e88} is satisfied by an arbitrary number of distinct species of particles $N$. It is interesting to remark that in our solution, on comparison to that presented in~\cite{BJP2} through the coordinate Bethe {\it ansatz}, has the advantage that the derived $S$-matrix do not depend on the size of the particles and the associativity condition or Yang-Baxter relation \rf{e94} is easier to be verified since it is the same, independently of the particle's sizes. 

        The spectral parameters $\{ k_j \}$ are fixed by the cyclic property of the trace in \rf{e73}. For each spectral parameter $k_j$ ($j=1, \ldots ,n$) the commutations relation \rf{e74}, \rf{e76}, \rf{e77} and \rf{e79} applied $j$ times, enable us to move the operator $Y_{k_j}^{(c_j)}$ to the left
\bea
\label{e95}
&&\mbox{Tr} \left[ Y_{k_1}^{(c_1)} \cdots Y_{k_{j-1}}^{(c_{j-1})}Y_{k_j}^{(c_j)} \cdots Y_{k_n}^{(c_n)}E^{L-\sum_{j=1}^N n_j s_j+n} \Omega_P \right] = e^{ik_j(L - \sum_{j=1}^N n_j s_j+n)} \nonumber \\
&&\sum_{c'_{j-1},c''_{j-1}} \sum_{c'_{j-2},c''_{j-2}} \cdots \sum_{c'_{1},c''_{1}} \left\{ e^{iP(s''_1-1)} S_{c'_1, c''_1}^{c_1, c''_2} (k_1,k_j) \cdots S_{c'_{j-2}, c''_{j-2}}^{c_{j-2}, c''_{j-1}} (k_{j-2},k_j) S_{c'_{j-1}, c''_{j-1}}^{c_{j-1}, c_j} (k_{j-1},k_j) \right. \nonumber \\
&&\left. \mbox{Tr} \left[ Y_{k_1}^{(c'_1)} \cdots Y_{k_{j-1}}^{(c'_{j-1})}Y_{k_{j+1}}^{(c_{j+1})} \cdots Y_{k_n}^{(c_n)} Y_{k_j}^{(c''_1)}E^{L-\sum_{j=1}^N n_j s_j+n} \Omega_P \right] \right\}.
\eea
Moving the operator $Y_{k_j}^{(c''_1)}$ for more $n-j$ times to the left and using the identity
\beq
\label{e96}
\sum_{c''_{j},c''_{j+1}=1}^N S_{c'_j, c''_j}^{c_j, c''_{j+1}} (k_j,k_j) = 1,
\eeq
we can write
\bea
\label{e97}
&&\mbox{Tr} \left[ Y_{k_1}^{(c_1)} \cdots Y_{k_n}^{(c_n)}E^{L-\sum_{j=1}^N n_j s_j+n} \Omega_P \right] = e^{ik_j(L - \sum_{j=1}^N n_j s_j+n)} \nonumber \\
&&\sum_{c'_1, \ldots ,c'_n} < c_1, \ldots ,c_n| {\cal T} |c'_1, \ldots , c'_n> \mbox{Tr} \left[ Y_{k_1}^{(c'_1)} \cdots Y_{k_n}^{(c'_n)} E^{L-\sum_{j=1}^N n_j s_j+n} \Omega_P \right] 
\eea
where
\beq
\label{e98}
< \{ c \}| {\cal T} | \{ c' \} > = \sum_{c''_1, \cdots ,c''_n} \left\{ S_{c'_1, c''_1}^{c_1, c''_2} (k_1,k_j) \cdots S_{c'_j, c''_j}^{c_j, c''_{j+1}} (k_j,k_j) \cdots S_{c'_n, c''_n}^{c_n, c''_1} (k_n,k_j) e^{iP(s''_1-1)} \right\}.
\eeq
We identify in \rf{e98} ${\cal T}(k_j; \{ k_l \})$ as the $N^n \times N^n$-dimensional transfer matrix of an inhomogeneous vertex model (inhomogeneities $\{ k_l \}$) with Boltzmann weights given by \rf{e85} and \rf{e88}. The model is defined on a cylinder of perimeter $n$ with a seam along its axis producing the twisted boundary condition
\beq
\label{e99}
S_{c'_n, c''_n}^{c_n, c''_{n+1}} (k_n,k) = S_{c'_n, c''_n}^{c_n, c''_1} (k_n,k) e^{iP(s''_1-1)},
\eeq
where as always, $P$ is the momentum of the eigenstate.
The relation \rf{e98} give us the conditions for the spectral parameters
\beq
\label{e100}
e^{-ik_j(L + n - \sum_{i=1}^N n_i s_i)} = \Lambda (k_j, \{ k_l \}) \;\;\; j=1, \ldots ,n,
\eeq
where $\Lambda (k_j, \{ k_l \})$ are the eigenvalues of the transfer matrix \rf{e98}. The problem of fixing the spectral parameters $\{ k_j \}$ reduces to the evaluation of the eigenvalues of the transfer matrix \rf{e98}. This can be done straightforwardly through the coordinate Bethe {\it ansatz} as in~\cite{BJP2}. Extracting these eigenvalues from~\cite{BJP2} we obtain equations that coincide with the Bethe {\it ansatz} equations for this general model (see Eqs. (71)-(73) in~\cite{BJP2}).

        Let us consider the more general Hamiltonian given in \rf{e69}, where now each molecule have an arbitrary size, independently of the sizes of the other molecules belonging to its class. The solution of this problem was not derived through the coordinate Bethe {\it ansatz} since is not simple in that formulation. The Hamiltonian \rf{e69} is composed by block disjoint eigenvectors labeled by $\{ c_1,s_1; \ldots ;c_n,s_n \}$ ($c_j=1, \ldots, N; \; s_j=1,2, \ldots ;\; j=1, \ldots ,n$) that specifies the classes and sizes of each individual particle. An arbitrary eigenfunction of \rf{e69} is given by a generalization of \rf{e71}, namely
\beq
\label{e101}
|c_1,s_1; \ldots ;c_n,s_n \rangle = \sum_{\{ c,s \}} \sum_{\{ x \}^* } f(x_1, c_1, s_1; \ldots ; x_n, c_n, s_n) |x_1,c_1,s_1; \ldots ;x_n,c_n,s_n \rangle,
\eeq
where $|x_1,c_1,s_1; \ldots ;x_n,c_n,s_n \rangle$ denotes the configuration where the particle located at $x_i$ ($i=1, \ldots ,L$) belong to the class $c_i$ ($c_i=1, \ldots ,N$) and has size $s_i$ ($s_i=1,2, \ldots$). The summations $\{ c,s \}=\{ c_{p_1},s_{p_1}; \ldots ; c_{p_n},s_{p_n} \}$ extends over all the permutations of particles. The summation $\{ x \}^*=\{ x_1, \ldots ,x_n \}^*$ runs, for each permutation $\{ c,s \}$, in the set of non-decreasing integers satisfying
\bea
\label{e102}
x_{i_1} &\geq& x_i + s_i \;\;\; i=1, \ldots ,n-1 \nonumber \\
s_1 &\leq& x_n - x_1 \leq L - s_n.
\eea
Our matrix product {\it ansatz} asserts that the eigenfunctions with a given momentum $P$, have amplitudes
\bea
f(x_1, c_1, s_1; &&\ldots ; x_n, c_n, s_n) = \nonumber \\
 &&\mbox{Tr} \left[ E^{x_1-1}Y^{(c_1,s_1)}E^{x_2-x_1-1}Y^{(c_2,s_2)} \cdots E^{x_n-x_{n-1}-1}Y^{(c_n,s_n)}E^{L-x_n} \Omega_P \right], \nonumber 
\eea
where the matrices $E$ are associated to the empty sites and $Y^{(c_j,s_j)}$ to the sites occupied by particles of class $c_j$ and having a size $s_j$ ($j=1, \ldots ,n$). The matrices $\Omega_P$, as before, fix the momentum $P$ of the eigenstates
\beq
\label{e104}
E \Omega_P = e^{-iP} \Omega_P E, \;\;\; Y^{(c,s)} \Omega_P = e^{-iP} \Omega_P Y^{(c,s)}.
\eeq
The solution of this general problem follows, {\it mutatis mutandis}, the derivation we did in \rf{e75}-\rf{e100}. The energy and momentum of the eigenstate is given by \rf{e93} where the spectral parameters are introduced by the generalization of \rf{e90}:
\beq
\label{e105}
Y^{(c,s)} = \sum_{i=1}^n Y_{k_i}^{(c,s)} E^{2-s},
\eeq
where the spectral parameter matrices $Y_{k_j}^{(c,s)}$ satisfy the algebra
\beq
\label{e106}
E Y_{k_j}^{(c,s)} = e^{k_j} Y_{k_j}^{(c,s)} E \;\;\; (j=1, \ldots ,n; \;\; c=1,2, \ldots ,N; \;\; s=1,2, \ldots),
\eeq
and from \rf{e104} 
\beq
\label{e107}
Y_{k_j}^{(c,s)} \Omega_P = e^{iP(1-s)} \Omega_P Y_{k_j}^{(c,s)} \;\;\; (j=1, \ldots ,n; \;\; c=1,2, \ldots ,N; \;\; s=1,2, \ldots).
\eeq
The algebraic relations among the $\{ Y_{k_j}^{(c,s)} \}$ are given by the generalization of \rf{e87}, namely,
\bea
\label{e108}
Y_{k_l}^{(c_1,s_1)}Y_{k_m}^{(c_2,s_2)} &=& \sum_{c'_1, c'_2 = 1}^N S_{c'_1, c'_2}^{c_1, c_2}(k_l,k_m) Y_{k_m}^{(c'_2,s_1)} Y_{k_l}^{(c'_1,s_2)} \;\;\; (k_l \neq k_m) \nonumber \\
Y_{k_l}^{(c_1,s_1)}Y_{k_l}^{(c_2,s_2)} &=& 0,
\eea
where ($l,m=1, \ldots ,n$), $c_1,c_2=1, \ldots ,N$ and $S_{c'_1, c'_2}^{c_1, c_2}(k_l,k_m)$ are the same $S$-matrix defined in \rf{e85}, \rf{e88} and \rf{e89}. It is interesting to observe that the condition of existence of a single relation among the words $\cdots Y_{k_1}^{(\alpha,s_1)}Y_{k_2}^{(\beta,s_2)}Y_{k_3}^{(\gamma,s_3)} \cdots$ and $\cdots Y_{k_3}^{(\gamma,s_1)}Y_{k_2}^{(\beta,s_2)}Y_{k_3}^{(\alpha,s_3)} \cdots$ reproduces the Yang-Baxter relation \rf{e94}, as before.

         The spectral parameters $\{ k_j \}$, as before, are fixed by the cyclic property of the trace and the algebraic relations \rf{e104}-\rf{e108}. Using these relations we can move the operator $Y_{k_j}^{(c_j,s_j)}$ to the left as in \rf{e95}-\rf{e98},
\bea
\label{e109}
&&\mbox{Tr} \left[ Y_{k_1}^{(c_1,s_1)}Y_{k_2}^{(c_2,s_2)} \cdots Y_{k_n}^{(c_n,s_n)}E^{L-\sum_{j=1}^N (s_j-1)} \Omega_P \right] = e^{ik_j[L - \sum_{j=1}^N (s_j-1)]} e^{iP(s_1-1)} \nonumber \\
&&\sum_{c'_1, \ldots , c'_n} < c_1, \ldots ,c_n| \tilde{{\cal T}} | c'_1, \ldots ,c'_n > \mbox{Tr} \left[ Y_{k_1}^{(c'_1,s_2)}Y_{k_2}^{(c'_2,s_3)} \cdots Y_{k_n}^{(c'_n,s_1)}E^{L-\sum_{j=1}^N (s_j-1)} \Omega_P \right]
\eea
where now
\beq
\label{e110}
< \{ c \}| \tilde{{\cal T}} | \{ c' \} > = \sum_{c''_1, \cdots ,c''_n} \left\{ S_{c'_1, c''_1}^{c_1, c''_2} (k_1,k_j) \cdots S_{c'_n, c''_n}^{c_n, c''_1} (k_n,k_j) \right\},
\eeq
is different from \rf{e98} since now it corresponds to a transfer matrix of a vertex model in a cylinder of perimeter $n$ with no seam (periodic boundary condition). If we iterate $n-1$ times the procedure used in obtaining \rf{e109} we obtain:
\bea
\label{e111}
&&\mbox{Tr} \left[ Y_{k_1}^{(c_1,s_1)}Y_{k_2}^{(c_2,s_2)} \cdots Y_{k_n}^{(c_n,s_n)}E^{L-\sum_{j=1}^N (s_j-1)} \Omega_P \right] = e^{ink_j[L - \sum_{j=1}^N (s_j-1)]} e^{iP \sum_{j=1}^n(s_j-1)} \nonumber \\
&&\sum_{c'_1, \ldots , c'_n} < c_1, \ldots ,c_n| \tilde{{\cal T}}^n | c'_1, \ldots ,c'_n > \mbox{Tr} \left[ Y_{k_1}^{(c'_1,s_1)} \cdots Y_{k_n}^{(c'_n,s_n)}E^{L-\sum_{j=1}^N (s_j-1)} \Omega_P \right].
\eea
Consequently the spectral parameters $\{ k_j \}$ should satisfy
\beq
\label{e112}
e^{-ik_j(L + n - \sum_{i=1}^n s_i)} = e^{i\frac{2\pi}{n}r} e^{iP(\bar{s}-1)} \tilde{\Lambda} (k_j, \{ k_l \}) \;\;\; j=1, \ldots ,n; \;\;\;\ r=0,1, \ldots, n-1,
\eeq
where $\tilde{\Lambda} (k_j, \{ k_l \})$ is an eigenvalue of the transfer matrix $\tilde{\cal T}$ given in \rf{e110}, and $\bar{s}=\sum_{j=1}^n \frac{s_j}{n}$ is the average size of the particles. The eigenvalues $\tilde{\Lambda} (k_j, \{ k_l \})$ can be obtained from the diagonalization of \rf{e110} through the coordinate Bethe {\it ansatz} and they are given by Eqs. (67) and (70) with $\Phi_{\alpha} = 1$ of~\cite{BJP2}. We finally have the conditions that fix the spectral parameters of this general problem 
\bea
\label{e113}
&&e^{-ik_j[L+n-\sum_{j=1}^n s_j]} = (-1)^{n-1} e^{i\frac{2\pi}{n}r}  e^{iP(\bar{s}-1)} \prod_{j'=1\;(j' \neq j)}^n \frac{\epsilon_+ + \epsilon_- e^{i(k_j+k_{j'})} - e^{ik_j} }{\epsilon_+ + \epsilon_- e^{i(k_j+k_{j'})} - e^{ik_{j'}}} \nonumber \\
&&\times \prod_{l=1}^{m_1} \frac{\epsilon_+ (e^{ik_l^{(1)}} - e^{ik_j} )}{\epsilon_+ + \epsilon_- e^{i(k_l^{(1)}+k_j)} - e^{ik_j}}, \;\;\; j=1,2, \ldots , n,
\eea
where the auxiliary complex parameters $\{ k_j^{(l)}, \; l=0, \ldots, N-1; \; j=1, \ldots ,m_l \}$ are fixed by the equations
\bea
\label{e114}
\prod_{\beta=1}^{m_l} \frac{\epsilon_+ (e^{ik_{\alpha}^{(l+1)}} - e^{ik_{\beta}^{(l)} } )}{\epsilon_+ + \epsilon_- e^{i(k_{\alpha}^{(l+1)}+k_{\beta}^{(l)})} - e^{ik_{\beta}^{(l)}} } = (-1)^{m_{l+1}} \prod_{\delta=1}^{m_{l+2}} \frac{\epsilon_+ (e^{ik_{\delta}^{(l+2)}} - e^{ik_{\alpha}^{(l+1)} } )}{\epsilon_+ + \epsilon_- e^{i(k_{\delta}^{(l+2)}+k_{\alpha}^{(l+1)})} - e^{ik_{\alpha}^{(l+1)}} }\nonumber \\
\times \prod_{\alpha'=1 \;(\alpha' \neq \alpha)}^{m_{l+1}} \frac{\epsilon_+ + \epsilon_- e^{i(k_{\alpha}^{(l+1)} + k_{\alpha'}^{(l+1)})} - e^{ik_{\alpha}^{(l+1)}} }{\epsilon_+ + \epsilon_- e^{i(k_{\alpha}^{(l+1)} + k_{\alpha'}^{(l+1)})} - e^{ik_{\alpha'}^{(l+1)}}} \;\;\; l=0,1, \ldots , N-2; \;\;\; \alpha = 1, \ldots ,m_l,
\eea
where $n_j$ ($j=1, \ldots ,N$), as before, is the number of particles on class $j$ and $m_l = \sum_{j=1}^{N-l} n_j,\;\; l=0, \ldots , N; \; m_0 = n, m_N=0$, and $k_j^{(0)} = k_j$. The energies and momentum are given in terms of $\{ k_j \}$ by \rf{e93}. We can see from \rf{e113} and \rf{e114}, since for each value of $r$ ($r=0,1, \ldots ,n-1$) we have distinct solutions, the number of solutions is higher than that obtained previously. This should be expected since the particles now are completely distinguishable.

\section{ Conclusions and generalizations }

We have shown that all the exact results derived for the asymmetric exclusion problem and generalizations, through the coordinate Bethe {\it ansatz} can also be obtained in an elegant and unified view by an appropriate matrix product {\it ansatz}. According to this {\it ansatz} the amplitudes of the eigenfunctions of the associated Hamiltonian are given by traces of a product of matrices. The algebraic properties of the matrices appearing in the {\it ansatz} are fixed by the eigenvalue equation of the Hamiltonian. The existence of a well defined ratio among the several amplitudes of any eigenfunction imply the associativity of the algebra ruling the matrices defining the {\it ansatz}. In the case where we have more than a single kind of particles the condition of associativity of the algebra (see \rf{e94}) coincides with the Yang-Baxter relations~\cite{yang2,baxter}. Once the algebraic relations of the matrices are fixed the eigenfunctions we obtain coincides with those obtained through the coordinate Bethe {\it ansatz}. As an example see \rf{e42} for the case of diffusion of one kind of particle with a fixed size $s$.

         Differently from the Bethe {\it ansatz} solutions presented in~\cite{PRE},~\cite{BJP1} and~\cite{BJP2} the matrix product {\it ansatz} we formulate allow us to treat in a unified way the hard-core exclusion effects produced by the size of the particles. This virtue allowed a simple derivation of the quite complicated problem (see \rf{e113} and \rf{e114} in section $5$) where we have $N$ types of particles hierarchically ordered, but each particle being distinguishable and with a given specified size. The corresponding calculation through the coordinate Bethe {\it ansatz} is rater difficult. 

         The extension of the solution presented in section $5$ for the cases where the molecules are allowed to have a zero size is immediate and follows the same reasoning of sections $2$ and $3$. In the case of a single specie of molecule we can also extend our models allowing the molecules to have negative size ($s=-1,-2, \ldots$) as in~\cite{And1}. In this case since we do not have the interchange process, the particles have a well defined order on the lattice, apart of cyclic rotations, i. e., ($x_1 \leq x_2 \leq \cdots \leq x_n$). A particle $i$ with negative size $s$ allows a partial break of this ordering, i. e., ($x_1 \leq x_2 \leq \cdots \leq x_{i-1}-s \leq x_i \leq x_{i+1} \leq \cdots \leq x_n$).   

         We may also extend the matrix product {\it ansatz} presented in this paper to the cases where the lattice size has open ends~\cite{TB}. In those cases instead of the trace operation defining the amplitudes of the eigenfunction we have a single undefined matrix product, that can be fixed by a normalization of the corresponding eigenfunction.

         The success of our matrix product {\it ansatz} can also be tested~\cite{TB} on an enormous variety of known exactly integrable models, irrespective if the Hamiltonian is related or not to nonequilibrium stochastic models. We have shown that our matrix product {\it ansatz} can provide the exact solutions of the XXZ chain with arbitrary exclusion effects~\cite{alcbar3}, the Fateev-Zamolodchikov model~\cite{fateev}, the Izergin-Korepin model~\cite{izergin}, the $t$-$J$ model~\cite{schlo}, the Hubbard model~\cite{lieb} as well the generalized integrable models presented in~\cite{alcbar2} and~\cite{alcbar1}. In conclusion our results suggest the conjecture that all exact integrable model may have its eigenfunctions given by an appropriate matrix product {\it ansatz}.

\acknowledgements { This work has been partly supported by FAPESP, CNPq and CAPES (Brazilian agencies).  }
\newpage 
%


\newpage
\Large
\begin{center}
Figure Captions
\end{center}
\normalsize
\vspace{1cm}

\noindent Figure 1 - Example of configurations of molecules with distinct 
sizes $s$ in a lattice of size $L=5$. 

\end{document}